\documentclass[aps,prl,twocolumn,superscriptaddress,showpacs]{revtex4}
\usepackage{amssymb}
\usepackage{amsmath}
\usepackage{graphicx}
\usepackage{makeidx}
\usepackage{amsfonts}
\usepackage{bm}
\usepackage{color}

\input{tcilatex}

\begin{document}

\title{Giant Nernst-Ettingshausen Oscillations in Semiclassically Strong
Magnetic Fields}
\author{Igor A. Luk'yanchuk}
\affiliation{Laboratory of Condensed Matter Physics, University of Picardie Jules Verne,
Amiens, 80039, France}
\author{Andrei A. Varlamov}
\affiliation{CNR-SPIN, Viale del Politecnico 1, I-00133 Rome, Italy}
\author{Alexey V. Kavokin}
\affiliation{Physics and Astronomy School, University of
Southampton, Highfield, Southampton, SO171BJ, United Kingdom}
\date{\today }

\begin{abstract}
We consider the Nernst-Ettingshausen (NE) effect in the presence of
semiclassically strong magnetic fields for a quasi-two-dimensional
system with a parabolic or linear dispersion of carriers. We show
that the occurring giant oscillations of the NE coefficient are
coherent with the recent experimental observation in graphene,
graphite and bismuth. In the 2D case we find the exact shape of
these oscillations and show that their magnitude decreases/increases
with enhancement of the Fermi energy for Dirac fermions/normal
carriers. With a crossover to 3D spectrum the phase of oscillations
shifts, their amplitude decreases and the peaks become
asymmetric\textbf{.}
\end{abstract}

\maketitle

The Nernst-Ettingshausen (NE) effect in metals \cite{1886Ettingshausen} is a
thermoelectric counterpart of the Hall effect. The effect consists in
induction of an electric field $E_{y}$ normal to the mutually perpendicular
magnetic field $H$ ($\parallel z$) and temperature gradient $\nabla _{x}T$.
All electric circuits are supposed to be broken: $J_{x}=J_{y}=0$ and heat
flow along y-axis to be absent (adiabatic conditions). Quantitatively,\ the
effect is characterized by the NE coefficient.%
\begin{equation*}
\nu =\frac{E_{y}}{(-\nabla _{x}T)H}.
\end{equation*}%
The NE coefficient varies by several orders of magnitude in different
materials ranging from about $7mV\cdot K^{-1}T^{-1}$ in bismuth\textbf{\ }up
to $10^{-5}mV\cdot K^{-1}T^{-1}$ in some metals \cite{2007Behnia}.

The NE effect was discovered in 1886 and remained poorly understood
until 1948 when Sondheimer \cite{1948Sondheimer}, using the
classical Mott formula for the thermoconductivity tensor, calculated
$\nu $ for a degenerated electron system. It has been linked to the
energy derivative of the Hall angle $\theta =\sigma _{xy}/\sigma
_{xx}$. Within this model, $\nu $ was found to be independent on the
magnetic field in weak fields and to decrease as $H^{-2}$ in the
region of semiclassically strong fields, where the
cyclotron frequency $\omega _{c}$ is larger than the\ inverse scattering time%
\textbf{\ }$\tau ^{-1}$. In 1964, Obraztsov \cite{1964Obraztsov}
suggested that  magnetization currents (i.e. electric currents
induced due to inhomogeneous distribution of magnetization in the
sample) can contribute supplementary to the NE effect.

The giant oscillations of $\nu $ were firstly experimentally observed in
1959 in zinc by Bergeron \textit{et al} \cite{1959Bergeron} who
qualitatively ascribed the phenomenon to crossing of the electronic Fermi
energy by Landau levels (LL). Similarly to de Haas - van Alphen (dHvA)
oscillations of magnetization and Shubnikov - de Haas (SdH) oscillations of
conductivity, in the NE oscillations the corresponding\textbf{\ }quantizing
fields are given by Lifshitz-Onsager condition \cite{1955Lifshitz}:
\begin{equation}
S\left( \mu \right) =\left( k+\gamma _{\sigma }\right) 2\pi \hbar \frac{%
eH_{k\sigma }}{c},  \label{LO}
\end{equation}%
where $S\left( \mu \right) $ is the cross section of Fermi surface (FS) of
the orbital electron motion at $p_{z}=0$, $\mu $ is the chemical potential, $%
k$ is integer. Here $\gamma _{\sigma }=\gamma +\frac{1}{2}\frac{m^{\ast }}{m}%
\sigma $ with $\sigma =\pm 1,$ and\ the electron cyclotron mass $m^{\ast }=%
\frac{1}{2\pi }\frac{dS}{d\mu }$ \cite{1955Lifshitz}.

Very recently, the NE effect has been measured \cite{2009Zuev,2009Checkelsky}
and theoretically analyzed \cite{2010Bergman} in graphene. Surprisingly, it
has been found that $\nu $ changes its sign at quantizing field in graphene
while it has maxima in zinc \cite{1959Bergeron} and bismuth \cite%
{2007aBehnia}. Zhu \textit{et al}. \cite{2009Zhu} demonstrated that such
untypical behavior of $\nu (H)$ observed in graphene is not reproduced in
graphite. They concluded that piling of multiple graphene layers leads to a
topological phase transition in the spectrum of charge carriers, so that
graphite behaves as a 3D crystal despite of its apparent structural
anisotropy and of\textit{\ }similarity of its electronic properties to those
of graphene.

Another challenging property of quantum oscillations is the possibility to
distinguish between two types of charge carriers, having the topologically
different parameter $\gamma $ \cite{Falkovsky,1999Mikitik}: $\gamma =\frac{1%
}{2}$ for the normal carriers (NC) with parabolic 2D dispersion and linear
LL quantization:
\begin{equation*}
\text{NC:}\quad \varepsilon (p{_{\perp }})=\frac{{p_{\perp }^{2}}}{{%
2m_{\perp }}},\quad \varepsilon _{k}=2\mu _{B}H\frac{m}{m_{\bot }}\left( k+%
\frac{1}{2}\right) ,
\end{equation*}%
and $\gamma =0$ for the Dirac fermions (DF) having the linear two-branch
spectrum and $\sim k^{1/2}$ LL quantization:
\begin{equation*}
\text{DF:}\quad \quad \varepsilon (p{_{\perp }})=\pm v|{p_{\perp }|,}\quad
\varepsilon _{k}=\pm \left[ 4mv^{2}\mu _{B}H~k\right] ^{1/2},
\end{equation*}%
${p_{\perp }}$ and $m_{\bot }$ being momentum and effective mass in the
plane normal to the magnetic field, $m$ is the free electron mass, $v$ is
the Fermi velocity and $\mu _{B}=e\hslash /2mc$ is the Bohr magneton.

In this Letter we propose a simple thermodynamic approach to the description
of the NE effect which allows linking the oscillations of the NE coefficient
to the oscillations of the magnetization. Both thermal (Sondheimer) and
magnetization (Obraztsov) contributions to the Nernst coefficient are
evaluated analytically for a quasi-two dimensional (q2D) electronic system
with either parabolic or Dirac spectrum. In the 2D limit for the Dirac
spectrum we recover the behavior of the NE coefficient observed in graphene
\cite{2009Zuev,2009Checkelsky} while the recent data of Zhu et al. \cite%
{2009Zhu} on graphite correspond to the 3D limit.

\emph{Thermodynamic approach. }The NE coefficient is measured in the absence
of the electric current flowing through the system along the temperature
gradient. This is why the system can be assumed to be in thermodynamic
equilibrium where the electrochemical potential $\mu +e\varphi =\mathrm{const%
}$, with $\varphi $ being the electrostatic potential. Hence the effect of
the temperature gradient is reduced to the appearance of an effective
electric field in the $x$- direction $E_{x}=\nabla _{x}\mu /e$. In this way,
the problem is reduced to the classical Hall problem, which allows us to
obtain the thermal contribution to the NE coefficient:
\begin{equation}
\nu ^{\mathrm{therm}}=\frac{\sigma _{xx}}{e^{2}nc}\left( \frac{d\mu }{dT}%
\right) ,  \label{2}
\end{equation}%
where $\sigma _{xx}$ is the diagonal component of the conductivity
tensor, $n $ is the concentration of carriers. This simple formula
reproduces Sondheimer's result for a normal metal, fluctuation
contribution to the NE coefficient in a superconductor above
$T_{c}$, etc. \cite{2009Varlamov,2009Serbyn}.

The additional contribution to the NE coefficient appearing due to the
spatial dependence of magnetization in the sample can be found from the
Ampere law. The magnetization current density is $\mathbf{j}^{\mathrm{mag}}=%
\frac{c}{4\pi }\nabla \times \mathbf{B},$where $\mathbf{B}=\mathbf{H}+4\pi
\mathbf{M}$, $\mathbf{H}$ is the spatially homogeneous external magnetic
field, $\mathbf{M}$ is the magnetization, which can be temperature and,
henceforth, coordinate dependent. In the case under consideration one can
express the magnetization current as $j_{y}^{\mathrm{mag}}=-c\left(
dM/dT\right) \nabla _{x}T$ \ \cite{1964Obraztsov} and the corresponding
contribution to the electric field in the $y$- direction (Nernst field) as $%
E_{y}^{\mathrm{mag}}=\rho _{yy}j_{y}^{\mathrm{mag}}$ , where $\rho _{yy}$ is
the diagonal component of the resistivity tensor ($\rho _{yy}=\rho _{xx}$).
The magnetization contribution to the NE coefficient reads as%
\begin{equation}
\nu ^{\mathrm{mag}}=\frac{c\rho _{yy}}{H}\left( \frac{dM}{dT}\right) .
\label{5}
\end{equation}

The Eqs. (\ref{2}) and (\ref{5}) reveal the essential physics of Nernst
oscillations in the quantizing magnetic fields. In particular, one can see
that the NE coefficient is dependent on the diagonal components of
conductivity and resistivity tensors. Their oscillations as a function of
the magnetic field constitute the SdH effect. The giant Nernst oscillations
have been observed even in the regime where the SdH effect is weak in
graphene (at $H<3T$) \cite{2009Checkelsky} and in graphite \cite{2009Zhu}.
This is why one should attribute the giant NE coefficient oscillations to
the remaining factors in the Eqs. (\ref{2}) and (\ref{5}), namely, to the
temperature derivatives of the chemical potential and magnetization, $d\mu
/dT$ and $dM/dT$, respectively. Remarkably, to evaluate these quantities no
supplementary knowledge of the transport properties of the system is needed.
These derivatives can be expressed in terms of the thermodynamic potential
of the system:
\begin{equation}
\frac{d\mu }{dT}=\frac{\partial ^{2}\Omega }{\partial T\partial \mu }\left(
\frac{\partial ^{2}\Omega }{\partial \mu ^{2}}\right) _{T}^{-1},\qquad \frac{%
dM}{dT}=\frac{\partial ^{2}\Omega }{\partial T\partial H}.  \label{6}
\end{equation}%
To be more specific, we consider the quasi-2D system with the dispersion
\begin{equation}
\varepsilon (p_{\perp },p_{z})=\varepsilon _{\perp }(p_{\perp })+2t\sin
\frac{p_{z}}{\hbar }d.  \label{disp}
\end{equation}%
This model allows us to describe the 2D-3D dimensional crossover by
variation of the hopping parameter $t$ from $t_{2D}=0$ to $t_{3D}\sim
\varepsilon _{F}$. The corresponding expression for the oscillating part of $%
\Omega $ (denoted by tilde), derived by Champel and Mineev for the parabolic
dispersion \cite{2001Champel} (see also \cite{2002Bratkovsky}) and
generalized in \cite{2004Lukyanchuk} for the arbitrary $\varepsilon _{\perp
}(p_{\perp })$ reads:
\begin{equation}
\widetilde{\Omega }=\frac{m^{\ast }}{2\pi \hbar ^{2}}\frac{\hbar ^{2}\omega
_{c}^{2}}{\pi ^{2}}\frac{1}{2}\sum_{l=1,\sigma =\pm 1}^{\infty }\frac{\psi
(\lambda l)}{l^{2}}\mathrm{Re}\Phi _{l\sigma }\left( \mu ,H\right) ,
\label{Omega}
\end{equation}%
with $\psi (\lambda l)=\frac{\lambda l}{\sinh \lambda l}$ and
\begin{equation}
\Phi _{l\sigma }\left( \mu ,H\right) =J_{0}\left( 2\pi l\frac{2t}{\hbar
\omega _{c}}\right) e^{\left[ -\frac{\Gamma }{\hbar \omega _{c}}+i\left(
\frac{c}{e\hbar }\frac{S(\mu )}{2\pi H}-\gamma _{\sigma }\right) \right]
2\pi l}.  \label{Phi}
\end{equation}%
Here $k_{B}=1$, $\lambda =2\frac{\pi ^{2}T}{\hbar \omega _{c}}$, $\Gamma $
is the Dingle LL broadening and $J_{0}$ is the Bessel function.  We present Eq. (\ref%
{Omega}) in the most general form using the parameters $S(\mu )$ at $p_{z}=0$%
, $m^{\ast },\omega _{c}$ and $\gamma _{\sigma }$. For NC $S=2\pi m_{\bot
}\mu $, $\ m^{\ast }=m_{\bot }$, $\omega _{c}=\frac{eH}{m_{\bot }c}$ and $%
\gamma _{\sigma }=\frac{1}{2}+\frac{1}{2}\frac{m_{\bot }}{m}\sigma $; for DF
$S=\pi \frac{\mu ^{2}}{v^{2}},\quad m^{\ast }=\frac{\mu }{v^{2}}$, $\omega
_{c}=\frac{eHv^{2}}{\mu c}$ and $\gamma _{\sigma }=\frac{1}{2}\frac{\mu }{%
mv^{2}}\sigma $. In the present derivation we assume a Lorentzian broadening
of Landau levels with a constant $\Gamma $. Such approximation can be
justified for $\omega _{c}\ll \varepsilon _{F}$ in the case of 3D system. In
2D systems it is \ expected to be valid only in the low field regime $\omega
_{c}\lesssim \tau ^{-1}.$ The oscillating parts of the chemical potential
and magnetization can be expressed using Eq. (\ref{6}) as:
\begin{eqnarray}
\frac{d\widetilde{\mu }}{dT}=-\frac{\mathrm{Im\,}\Xi ^{\left\{ 1\right\} }}{%
1+2\,\mathrm{Re\,}\Xi ^{\left\{ 0\right\} }},\frac{d\widetilde{M}}{dT}=\frac{%
n}{H}\frac{d\widetilde{\mu }}{dT},  \label{dmdT}
\\
\Xi ^{\left\{ \alpha \right\} }=\frac{1}{2}\sum_{l=1,\sigma =\pm
1}^{\infty }\psi ^{\left( \alpha \right) }\left( \lambda l\right)
\Phi _{l\sigma }\left( \varepsilon _{F},H\right)   \label{Theta}
\end{eqnarray}%
and $\psi ^{\left( \alpha \right) }\left( x\right) $ is the derivative of
the order of $\alpha =0,1$ of the function $\psi $. One can see from Eqs. (%
\ref{5}) and (\ref{dmdT}) that the NE\ coefficient oscillates proportionally
to the derivative of magnetization over temperature. This shows an important
link between NE and dHvA oscillations, which is universal and independent on
the dimensionality of the system and of the type of carriers.

It is convenient to express the NE coefficient as
\begin{equation}
\nu =\nu ^{\mathrm{therm}}+\nu ^{\mathrm{mag}}=\nu _{0}\left( H\right) +%
\widetilde{\nu }\left( H\right)   \label{nusum}
\end{equation}%
with $\nu _{0}\left( H\right) \ $\ and $\widetilde{\nu }\left( H\right) $
being the background and oscillating parts. The background part can be
evaluated in the Drude approximation as \cite{2009Varlamov}
\begin{equation}
\nu _{0}\left( H\right) =\frac{\pi ^{2}\tau }{6m^{\ast }c}\left( \frac{T}{%
\varepsilon _{F}}\right) \frac{1}{1+\left( \omega _{c}\tau \right) ^{2}}.
\label{kH}
\end{equation}%
The account for magnetization currents leads to the correction of the order
of $\left( \varepsilon _{F}\tau \right) ^{-2}$ with respect to Sondheimer
result described by Eq. (\ref{kH}).

The oscillating part of the Nernst coefficient can be written using Eqs. (%
\ref{2}),(\ref{5}) and (\ref{dmdT}) as:
\begin{equation}
\widetilde{\nu }\left( H\right) =-2\pi \kappa \left( H\right) \frac{\mathrm{%
Im\,}\Xi ^{\left\{ 1\right\} }}{1+2\,\mathrm{Re\,}\Xi ^{\left\{ 0\right\} }},
\label{tot}
\end{equation}%
with%
\begin{equation}
\kappa \left( H\right) =\frac{\sigma _{xx}(H)}{e^{2}nc}+\frac{cn\rho _{xx}(H)%
}{H^{2}}.  \label{kappaH}
\end{equation}%
\

In the Drude approximation for NC
\begin{equation}
\kappa _{\mathrm{Drude}}\left( H\right) =\frac{\tau }{m^{\ast }c}\frac{1}{%
(\omega _{c}\tau )^{2}}\frac{1+2(\omega _{c}\tau )^{2}}{1+(\omega _{c}\tau
)^{2}}.  \label{Drud}
\end{equation}

Equation (\ref{tot}) describes oscillations of the NE effect in the most
general form. It is valid for any type of the dispersion $\varepsilon
_{\perp }(p_{\perp })$ if $T,t$ $\ll \mu $.

\emph{The 2D case: graphene. }We start analysis\ of the Eq. (\ref{tot}) from
the pure 2D case where $t=0.$ In the low-temperature limit\textit{\ \ }$2\pi
^{2}T<\hbar \omega _{c}$ in Eq. (\ref{Omega}) $\lambda \ll 1$, hence $\psi
(\lambda l)\approx 1-\frac{1}{6}\lambda ^{2}l^{2}$. For $m^{\ast }<0.02m$
and $H=10T$ $\ $(typical in graphene experiments)$\ $this yields $T<10K$.
Since $m^{\ast }\ll m$ we neglect also the Zeeman splitting, assuming that $%
\gamma _{\sigma }=\gamma =0$ for NC and $\gamma _{\sigma }=\gamma =\frac{1}{2%
}$ for DF. The series $\Xi ^{\left\{ 0\right\} }$ and $\Xi ^{\left\{
1\right\} }$ in Eq. (\ref{tot}) in this case can be summed exactly which
gives:
\begin{equation}
\widetilde{\nu }^{(2D)}\left( \mu ,H\right) \!=\!\frac{2\pi ^{3}}{3}\!\frac{T%
}{\hbar \omega _{c}}\frac{\kappa \left( H\right) \,\,\sin 2\pi \left[ \frac{c%
}{e\hbar }\frac{S(\mu )}{2\pi H}-\gamma \right] }{\cosh \frac{2\pi \Gamma }{%
\hbar \omega _{c}}-\cos 2\pi \left[ \frac{c}{e\hbar }\frac{S(\mu )}{2\pi H}%
-\gamma \right] }.  \label{S1}
\end{equation}%
In the experimental configuration corresponding to the measurement of the NE
effect in graphene, the number of particles $n$ is fixed, so that \cite%
{2001Champel}:
\begin{equation}
n=-\left( \frac{\partial \Omega (\mu )}{\partial \mu }\right) _{H,T}=2\frac{%
S(\mu )}{\left( 2\pi \hbar \right) ^{2}}-\left( \frac{\partial \widetilde{%
\Omega }(\mu )}{\partial \mu }\right) _{H,T}=\mathrm{const}  \label{nS0}
\end{equation}%
(we assume the volume $V=1).$ This relation implicitly determines the
dependence of $\mu $ on $H,T$ for the given $n$. We note that the chemical
potential $\mu $ itself is a function of $H$ as follows from Eq. (\ref{nS0}%
), which in the 2D case can be written as:
\begin{equation*}
n=2\frac{S(\mu )}{\left( 2\pi \hbar \right) ^{2}}+\frac{m^{\ast }}{\hbar ^{2}%
}\frac{\hbar \omega _{c}}{\pi ^{2}}\arctan \frac{\sin 2\pi \left( \frac{c}{%
e\hbar }\frac{S(\mu )}{2\pi H}-\gamma \right) }{e^{\frac{2\pi \Gamma }{\hbar
\omega _{c}}}-\cos 2\pi \left( \frac{c}{e\hbar }\frac{S(\mu )}{2\pi H}%
-\gamma \right) }.
\end{equation*}%
\begin{figure}[t]
\centering \includegraphics [width=8cm] {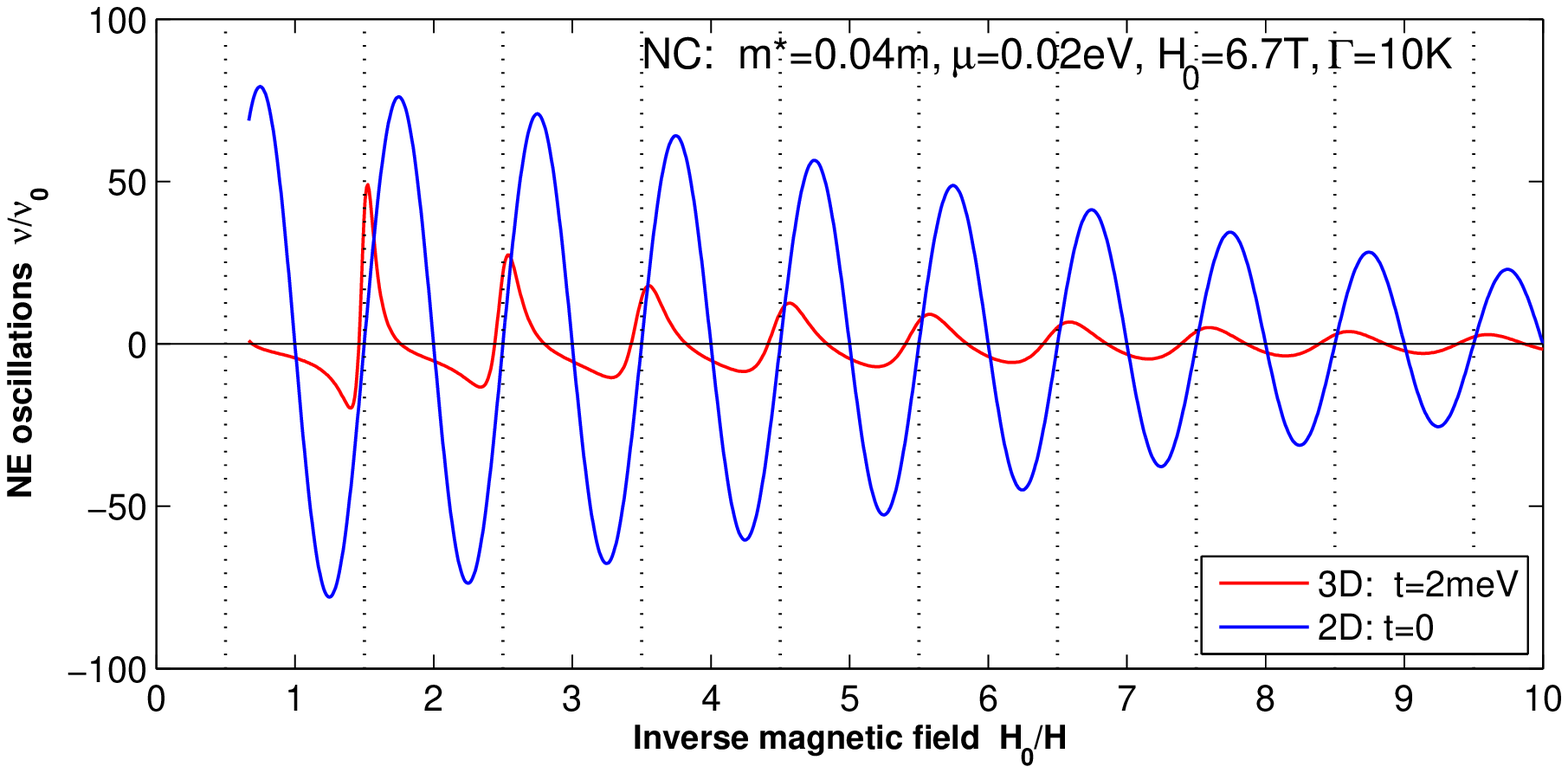}
\includegraphics
[width=4cm] {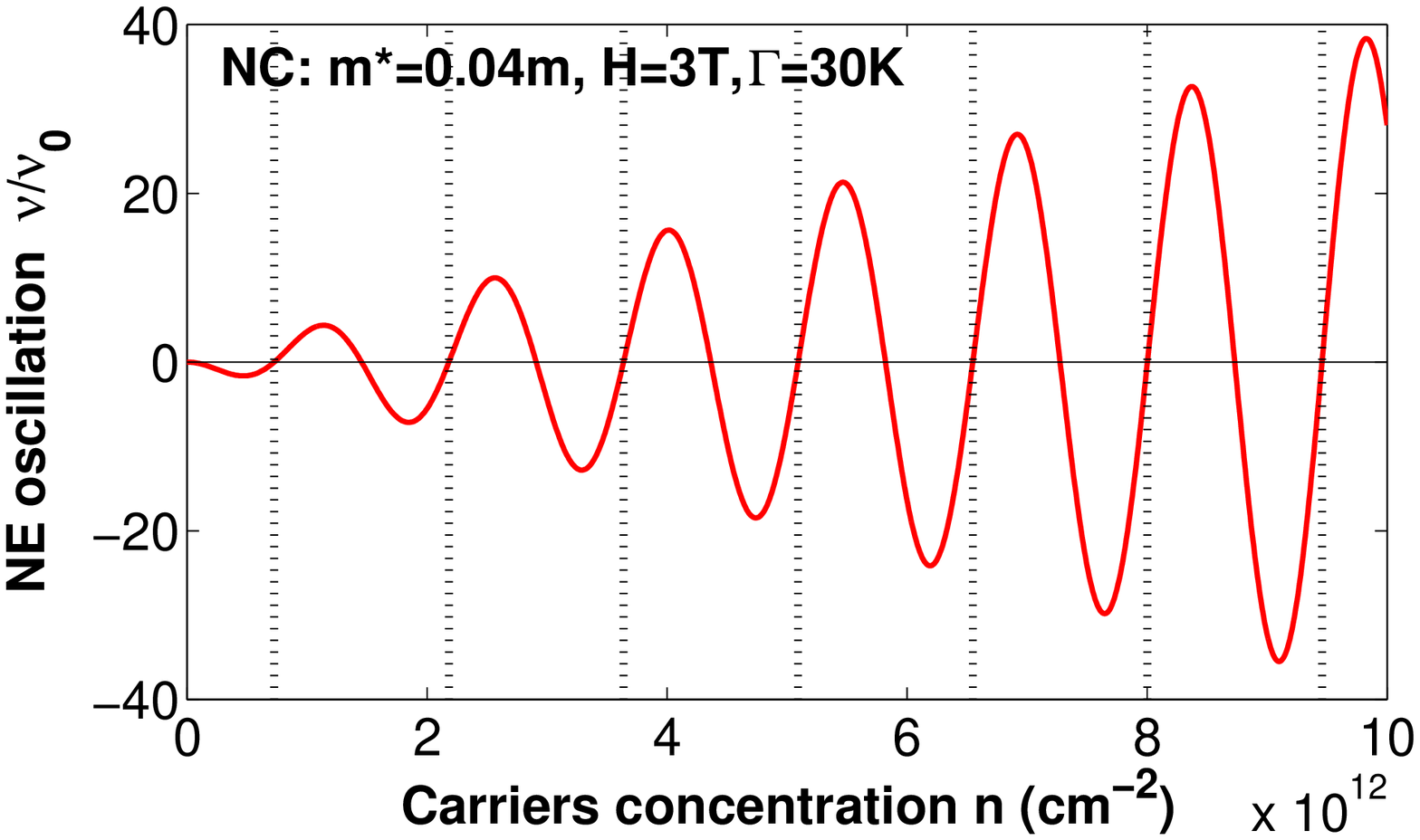} \centering
\includegraphics [width=4cm]{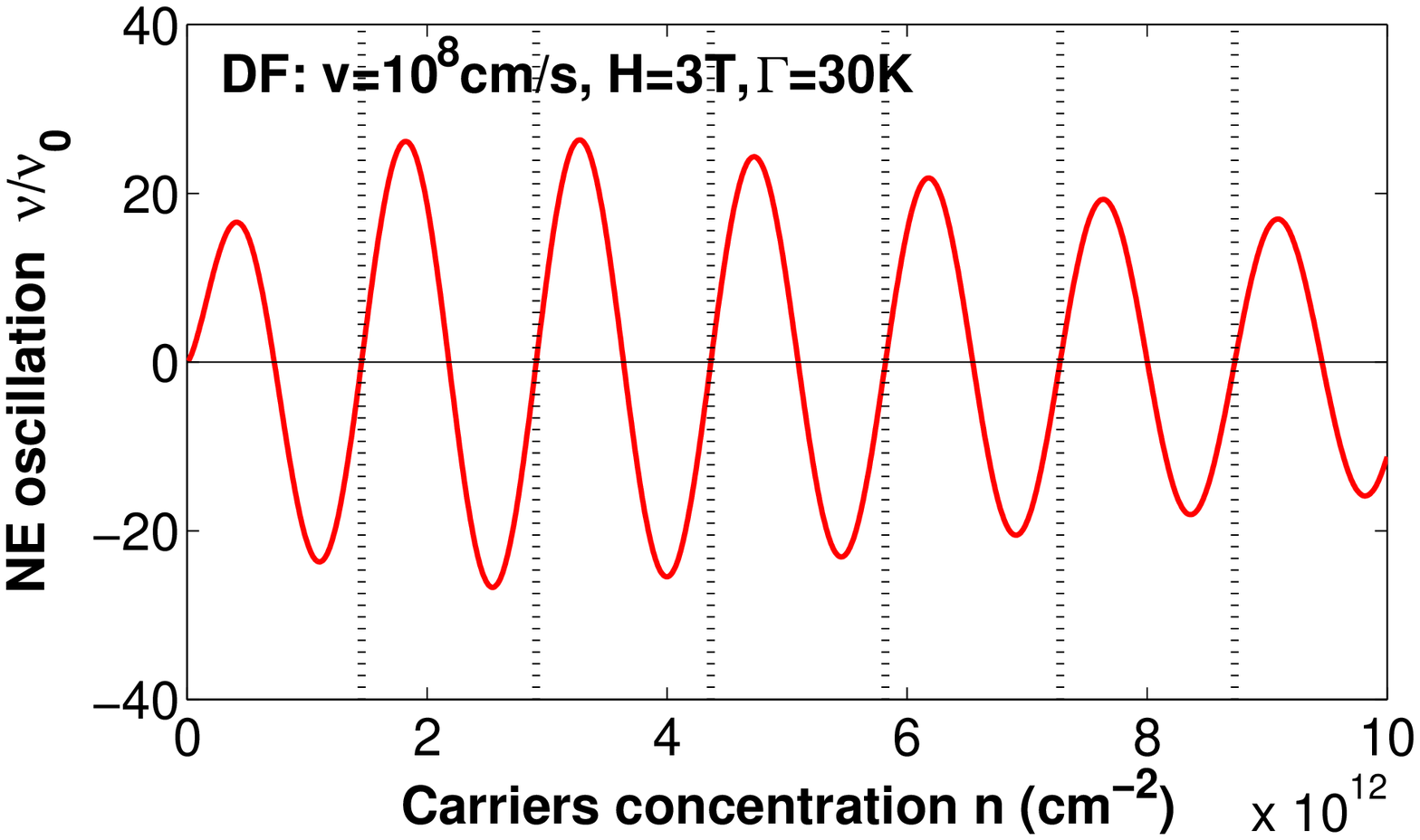}
\caption{Normalized Nernst-Ettingshausen (NE) oscillation as
function of the inverse magnetic field and carriers concentration
for normal carriers (NC) and Dirac fermions (DF). Dependence
$\protect\nu (H^{-1})$ for DF has the same profile as for NC but
shifted on half period. Vertical lines shows the quantization
condition (\protect\ref{LO}).} \label{FigU}
\end{figure}
This equation can be inverted for $S(\mu ):$
\begin{equation}
\frac{c}{e\hbar }\frac{S(\mu )}{2H}=\pi ^{2}\frac{\hbar c}{e}\frac{n}{H}%
-\arctan \frac{\sin 2\pi \left( \pi \frac{\hbar c}{e}\frac{n}{H}-\gamma
\right) }{e^{\frac{2\pi \Gamma }{\hbar \omega _{c}}}+\cos 2\pi \left( \pi
\frac{\hbar c}{e}\frac{n}{H}-\gamma \right) }.  \label{Smu}
\end{equation}%
Equation (\ref{Smu}) yields the dependence $\mu (n,H)$. Substituting it to
Eq. (\ref{S1}) after some cumbersome algebra one can find the oscillating
part of the Nernst coefficient explicitly:
\begin{equation}
\widetilde{\nu }^{\left( 2D\right) }\left( n,H\right) =\frac{2\pi ^{3}}{3}%
\frac{T}{\hbar \omega _{c}}\frac{\kappa \left( H\right) }{\sinh \frac{2\pi
\Gamma }{\hbar \omega _{c}}}\sin 2\pi \left( \pi \frac{\hbar c}{e}\frac{n}{H}%
-\gamma \right) ,  \label{N2Dn}
\end{equation}%
that is a strongly oscillating function. It crosses zero at the
intersections of LL and chemical potential, given by the condition $%
H=H_{k\sigma }$ defined by (\ref{LO}). The field depended factor $\kappa
\left( H\right) $\ is governed by magnetoresistance and is given by Eq.(\ref%
{kappaH}). At $\omega _{c}\tau \leq 1$\ where SdH oscillations are small, $%
\kappa \left( H\right) $\ \ can be roughly estimated using the Drude
approximation (\ref{Drud}). In particular, approaching the limit $\omega
_{c}\tau \sim 1$\ and assuming $\Gamma \sim \hbar /2\tau $ we obtain that $%
\kappa \left( H\right) \sim \frac{\tau }{m^{\ast }c}$\ and the amplitude of
NE oscillations\ is giant in comparison with the background: $\widetilde{\nu
}^{\left( 2D\right) }\sim \frac{\varepsilon _{F}}{\hbar \omega _{c}}\nu _{0}$%
. At higher fields $\omega _{c}\tau >1$, in the quantum Hall regime,
the shape of oscillations of the NE coefficient is affected by
strong variation of the magnetoresistance and Dingle temperature.
This can be taken into account by substitution of the field
dependent magnetoresistance and Dingle temperature into Eqs.
(\ref{tot}),(\ref{kappaH}).

The given by Eq. (\ref{N2Dn}) profiles of 2D\ NE oscillation as function of $%
H$ and $n$ for DF and NC\ are presented in Fig.1. Both our theory for DF and
experiment in graphene \cite{2009Zuev,2009Checkelsky} show a $\sin $-like
profile of the signal whose amplitude slightly decreases with increasing $n$%
. This tendency contradicts to the earlier theoretical predictions of the
classical Mott formula \cite{2009Zuev} that has been derived for a Boltzmann
gas of electrons. In contrast, the amplitude of NE oscillations increases
with increasing $n$ for the NC\ in a qualitative agreement with the Mott
formula.\

\emph{Quasi-2D and 3D cases. }In order to describe the NE effect in the
general quasi-2D case where $t\neq 0$ the Bessel function in the Eq. (\ref%
{Phi}) should be taken into account. The sums (\ref{Theta}) can be reduced
to the integrals by means of the Poisson transformation. Then integration
can be done analytically resulting in%
\begin{eqnarray}
\Xi ^{\left\{ 0\right\} }=\frac{1}{2}\sum_{\substack{ k=-\infty  \\
\sigma =\pm 1}}^{\infty }\frac{1}{2\pi \left[ \delta _{k\sigma
}^{2}\left( H\right) +\frac{4t^{2}}{\hbar ^{2}\omega
_{c}^{2}}\right] ^{1/2}}-\frac{1}{2}, \label{Sig0} \\
\Xi ^{\left\{ 1\right\} }=-\frac{1}{6}\frac{T}{\hbar \omega _{c}}\frac{1}{2}%
\sum_{\substack{ k=-\infty  \\ \sigma =\pm 1}}^{\infty }\frac{\delta
_{k\sigma }\left( H\right) }{\left[ \delta _{k\sigma }^{2}\left( H\right) +%
\frac{4t^{2}}{\hbar ^{2}\omega _{c}^{2}}\right] ^{3/2}},  \label{Sig1}
\end{eqnarray}%
where $\delta _{k\sigma }\left( H\right) =\frac{\Gamma }{\hbar \omega _{c}}-i%
\frac{c}{\hbar e}\frac{S}{2\pi }\left( H^{-1}-H_{k\sigma }^{-1}\right) $.
The NE coefficient is obtained by substitution of the Eqs. (\ref{Sig0}), and
(\ref{Sig1}) to Eq. (\ref{tot}). Resonances at $i\delta _{k\sigma }\left(
H\right) =\pm \frac{2t}{\hbar \omega _{c}}$ in $\widetilde{\nu }\left(
H\right) $ appear when the chemical potential crosses the quantized slices
of maximal (minimal) cross sections of the corrugated cylinder FS $S_{\max
(\min )}=S\pm 4\pi tm^{\ast }$.

In the wide quasi 2D interval $t<\left( \hbar \omega _{c}\right) ^{2}/\Gamma
$ the behavior of $\widetilde{\nu }^{(q2D)}\left( H\right) $\ close to $%
H=H_{k\sigma }$\ can be studied selecting in (\ref{Sig0}) and (\ref{Sig1})
only the resonant terms. With growth of $t$\ the positions of zeros shift
from $\mathrm{Im}\delta _{k\sigma }\left( H\right) =0$\ to $\mathrm{Im}%
\delta _{k\sigma }\left( H\right) =\pm \frac{2t}{\hbar \omega _{c}}$. The
superposition of two (for $S_{\max }$\ and $S_{\min }$) series of resonances
leads to the beats in $\widetilde{\nu }\left( H\right) $\ oscillations.

In the 3D limit $t>\left( \hbar \omega _{c}\right) ^{2}/\Gamma $ , $\mathrm{%
Re}\Xi ^{\left\{ 0\right\} }\ll 1$, so that $\Xi ^{\left\{ 0\right\} }$ can
be neglected in the denominator of Eq. (\ref{tot}). In the vicinity of $%
H=H_{k\sigma }$ one finds
\begin{equation}
\widetilde{\nu }^{(3D)}\left( H\right) =\mp \frac{\pi }{12}\frac{T\kappa
\left( H\right) }{\left( t\hbar \omega _{c}\right) ^{1/2}}\mathrm{Re\,}\frac{%
1}{\left[ \frac{2t}{\hbar \omega _{c}}\pm i\delta _{k}\left( H\right) \right]
^{3/2}},  \label{PEAK}
\end{equation}%
We assumed here the constant $\mu $ and neglected Zeeman splitting, taking $%
\delta _{k,\pm 1}=\delta _{k}$. The resonances in $\widetilde{\nu }\left(
H\right) $ described by Eq. (\ref{PEAK}) have the form of asymmetric spikes
with $\left\vert \widetilde{\nu }^{(3D)}\right\vert _{\max }/\left\vert
\widetilde{\nu }^{(3D)}\right\vert _{\min }\simeq 3.4$ as shown in Fig.\ref%
{FigU}. In the Drude approximation, the amplitude
\begin{equation}
\left\vert \widetilde{\nu }^{(3D)}\right\vert _{\max }\simeq 0.29\frac{%
\varepsilon _{F}}{\Gamma }\frac{\hbar \omega _{c}}{\left( t\Gamma \right)
^{1/2}}\nu _{0}\left( H\right)
\end{equation}%
is giant if $\frac{\varepsilon _{F}}{\Gamma }\frac{\hbar \omega _{c}}{\left(
t\Gamma \right) ^{1/2}}>1.$ \textbf{\ }

For 2D systems our calculations are valid for\ magnetic fields $\omega
_{c}\lesssim \tau ^{-1}$\ where one can neglect the quantum Hall
oscillations of conductivity. At higher fields the approach of Girvin and
Jonson \cite{1982Girvin}, based on the generalized Mott formula for the
thermopower tensor for 2D systems, seems to be more relevant. In 3D case the
range of applicability of our theory is given by $\omega _{c}\ll \varepsilon
_{F}$. Recently Bergman and Oganesyan \cite{2010Bergman} extended the
approach of Ref. \cite{1982Girvin} to calculate the off-diagonal
thermoelectric conductivity $\alpha _{xy}$\ for a 3D system at $\omega
_{c}\sim \varepsilon _{F}$. Although $\alpha _{xy}$\ constitute only the
part of the NE coefficient $\nu =-\left( \rho _{xx}\alpha _{xy}+\rho
_{xy}\alpha _{yy}\right) /H$, they reproduce quite well the measured in
graphite \cite{2009Zhu} sawtooth dependence of $\nu (H)$, having the
characteristic $(H_{k}-H)^{-\frac{1}{2}}$\ divergencies at resonances.

\emph{\ In conclusion}, we have obtained an analytical expression
for the oscillating NE constant in a 2D system with an arbitrary
electron dispersion, describing the recent experimental results in
graphene and predicting a qualitative difference in the NE
oscillations for NC and DF. We show that the giant oscillations of
the NE coefficient predicted and observed in a 2D case (graphene)
decrease significantly as the spectrum acquires a 3D character
(graphite). We describe analytically the shape of NE oscillations.
The NE oscillations are proportional to the temperature derivative
of the dHvA oscillations.

This work was supported by FP7-IRSES programs: ROBOCON and SIMTECH.

\end{document}